\documentclass[aps,prb,reprint,floatfix,longbibliography,twocolumn]{revtex4-2}
\usepackage{amsmath,amssymb,amsthm,hyperref,graphicx}
\hypersetup{hidelinks}
\usepackage[capitalise]{cleveref}
\usepackage{graphicx}
\usepackage{hyperref}
\hypersetup{colorlinks=true, linkcolor=blue, citecolor=red, urlcolor=blue}

\begin{document}
\title{Strong coupling between antiferromagnetic magnons and spoof surface plasmons}
\author{Yamin Sun$^{1,2}$}
\author{Ka Shen$^{2}$}
\author{H. Y. Yuan$^{1}$}
\email{hyyuan@zju.edu.cn}
\affiliation{$^{1}$Institute for Advanced Study in Physics, Zhejiang University, 310027 Hangzhou, China}
\affiliation{$^{2}$Center for Advanced Quantum Studies and School of Physics and Astronomy, Beijing Normal University, Beijing, China}

\date{\today}

\begin{abstract}
Hybrid magnonic system provides a versatile platform for coherent information exchange between spin excitations and other physical degrees of freedom. While strong coupling between magnons and spoof plasmons has been observed based on ferrimagnetic spheres and localized microwave resonators, it remains unexplored whether coherent magnon-plasmon coupling can be achieved in planar magnetic structures. Here we study the hybridization of antiferromagnetic (AFM) magnons and spoof surface plasmons in a planar heterostructure consisting of an AFM thin film, a dielectric spacer, and a structured metal surface. By analytically solving the coupled Maxwell and magnetization dynamics equation, we predict strong magnon-plasmon coupling in the terahertz regime, manifested by pronounced avoided crossings in the dispersion. The coupling originates from the spatial overlap between magnonic and plasmonic modes in the dielectric spacer, and can be efficiently tuned through geometric parameters of the system. The calculated cooperativity confirms that the hybrid system can operate in the strong coupling regime, enabling coherent information transfer between magnons and plasmons. Our results establish a planar platform for plasmon-magnon hybridization, which is more amenable to on-chip manipulation and integration.
\end{abstract}

\maketitle
\section{Introduction}
Hybrid magnonics manipulates the interplay of magnons with other physical excitations for information processing and has attracted significant attention recently \cite{YuanReview2022,BabakReview20221,LiReview2020,ZuoNJP2024}. Among the various physical excitations, including photons \cite{PhysRevLett.104.077202,PhysRevLett.111.127003}, phonons \cite{zhang2016cavity}, and qubits \cite{tabuchi2015coherent,lachance2020entanglement,XuPRL2023,WengNC2026}, strong magnon-photon coupling has been extensively studied in both the cavity and waveguide platforms and provides a versatile platform for coherent information transfer in hybrid information processing \cite{BabakReview20221}. More recently, the framework of hybrid magnonics was further extended to consider the interaction between magnons and surface plasmons \cite{bludov2019hybrid,costa2023strongly,yuan2025strong,dyrda2023magnonplasmon,xiong2024hybrid,yuan2024breaking,liu2024ghost,kuznetsov2025optical,qian2025unidirectional,PieterPRB2026,HiroPRB2026}, i.e. the collective excitation of itinerant electrons in metallic systems. Due to subwavelength confinement and strongly enhanced electromagnetic field, surface plasmons offer exciting opportunities in wave amplification, high-resolution imaging, and biosensing \cite{williaml.barnes2003surface,junxizhang2012surface}. However, whether and how we can harness the strong and localized electromagnetic field of surface plasmons to manipulate the spin degree of freedom is still an open and compelling question. 

A fundamental challenge in realizing magnon-plasmon coupling is the large frequency mismatch between conventional surface plasmons and magnons. Surface plasmons usually work at optical frequencies, whereas the magnon frequency ranges from gigahertz (GHz) to terahertz (THz) regime
\cite{kalinikos1986theory,v.baltz2018antiferromagnetic}, depending on the types of magnetic materials. This challenge can be partially overcome by employing plasmons in 2D materials, such as graphene, where electric gating and chemical doping allow the plasmon frequency to fall into the THz regime \cite{hwang2007dielectric,a.n.grigorenko2012graphene}. Theoretical studies predicted the coherent magnon-plasmon interaction in the hybrid structure of graphene and magnets \cite{bludov2019hybrid,costa2023strongly,yuan2025strong}. However, the fabrication of high-quality magnetic$|$graphene hybrid structures remains experimentally demanding, and thus the search for alternative platforms to study the magnon-plasmon interaction is necessary.

Spoof surface plasmons (SSPs) are electromagnetic surface waves in periodically structured metallic surfaces and provide a promising platform for extending plasmonics into the GHz and THz regimes \cite{pendry2004mimicking,garcia2022spoof}. SSPs exhibit dispersion and subwavelength field confinement analogous to conventional surface plasmons, while their frequencies can be flexibly engineered through geometric design using the mature nanofabrication technique \cite{pendry2004mimicking,hibbins2005experimental}. Recently, strong magnon-plasmon coupling has been experimentally demonstrated in hybrid systems consisting of magnetic spheres and localized SSP resonators  \cite{xiong2024hybrid,qian2025unidirectional}. An important open question is whether coherent magnon-plasmon coupling can also be realized for traveling-wave SSPs in planar geometries, which may be more attractive for on-chip integration and information transport. This motivates our current work.

In this work, we investigate the coupling between antiferromagnetic (AFM) magnons and traveling-wave spoof surface plasmons in a planar structure consisting of an AFM thin film, a dielectric spacer, and a metal grooved with one-dimensional periodic structures (Fig. \ref{fig:1}). By analytically solving Maxwell equations together with the Landau-Lifshitz-Gilbert (LLG) equations, We obtain the dispersion of the hybrid magnon-plasmon excitation with a clear avoided crossing structure between these two modes in the THz regime. The coupling strength can be conveniently  tuned by adjusting the geometric parameters of the grooved metals, which provides a practical advantage over 2D material-based tuning approaches. For widely used AFM materials (FeF$_2$, MnF$_2$ and NiO), the calculated cooperativity well exceeds unity, confirming that the hybrid system operates in the strong coupling regime. Our results establish a traveling-wave planar platform for coherent information transfer between magnons and plasmons, and may further stimulate the design of magnon-photon devices operating in the sub-THz and THz frequencies.

\section{Model and method}
\begin{figure}
  \centering
    \includegraphics[width=0.9\linewidth]{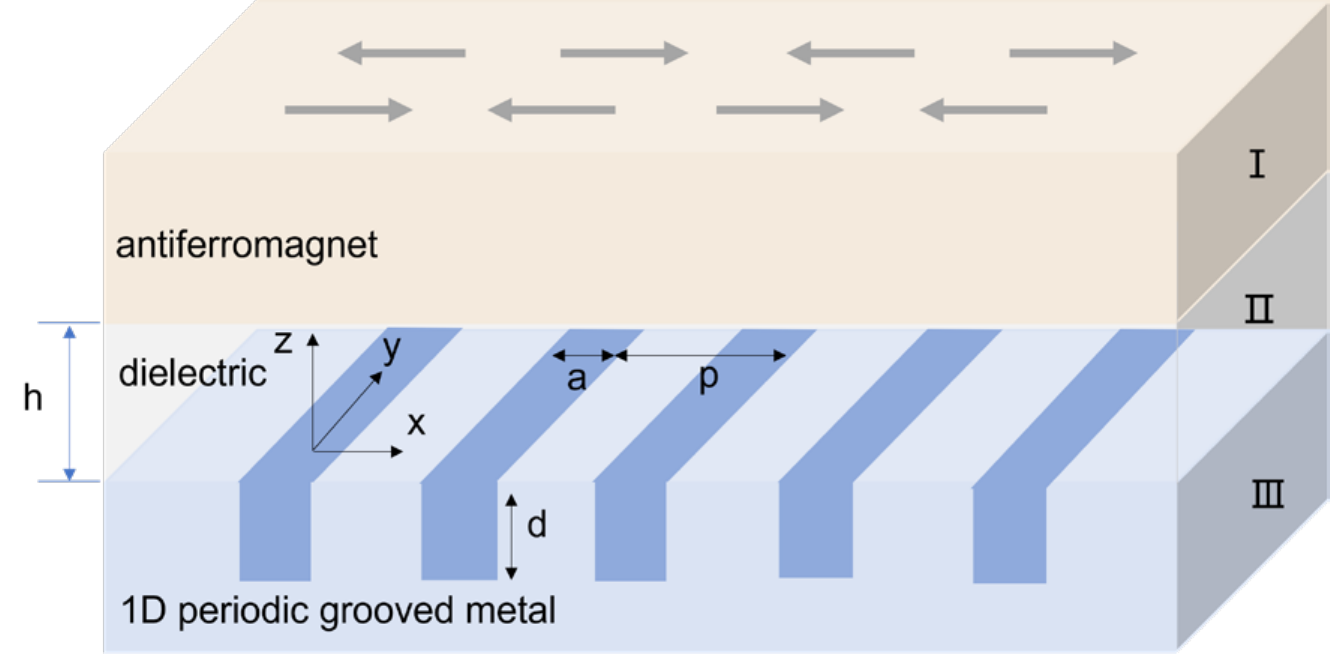}
    \caption{Schematic of the planar hybrid structure composed of an AFM thin film, dielectric spacer and one dimensional-periodic grooved metal. The classical ground state of the AFM layer is a Néel state with spins aligned antiparallel to each other along the $\pm x$ directions.}
    \label{fig:1}
\end{figure}

We consider a planar hybrid structure consisting of a two-sublattice AFM insulator, a dielectric spacer and a one dimensional (1D) periodically grooved metals, as shown in Fig. \ref{fig:1}. Region I is a uniaxial two-sublattice AFM insulator, with two equivalent magnetic sublattices whose magnetic moments are aligned antiparallel to each other. Such an AFM structure supports linearly polarized spin waves, which are well suited for coupling to transverse magnetic (TM) spoof surface plasmons, as we shall discuss in the following. Region II is a dielectric medium, where the evanescent magnetic fields associated with the spin waves and plasmons spatially overlap. Its thickness provides an effective means to tune the coupling strength between magnons and plasmons. Region III is a structured metal with 1D periodic grooves etched on its surface. The periodic corrugation supports the existence of bound electromagnetic modes, i.e. TM spoof surface plasmons, whose dispersion can be engineered in the THz range \cite{garcia2022spoof}.

The electromagnetic properties of this heterostructure are governed by the Maxwell equations
\begin{subequations}
    \begin{gather}
        \nabla  \times {{\bf{E}}} =  - \frac{\partial }{{\partial {\rm{t}}}}{{\bf{B}}}, \\
        \nabla  \times {{\bf{H}}} = \frac{\partial }{{\partial {\rm{t}}}}{{\bf{D}}},
    \end{gather}
    \label{eq2}
\end{subequations}
\hspace*{-0.5em}where ${{\bf{E}}}$  and  ${{\bf{H}}}$ are, respectively, the electric and magnetic fields in each region. Electric displacement and magnetic inductance are defined as ${{\bf{D}}} = \epsilon_0 \epsilon {{\bf{E}}}$  and  ${{\bf{B}}} = \mu_0 (\bf{H} + \bf{M}) \equiv \mu_0 \mu {{\bf{H}}}$, where $\mu$ and $\epsilon$ are the relative permeability and dielectric constant.
By eliminating the electric freedoms in Eq. \eqref{eq2}, we obtain the wave equation for the magnetic freedom in the three regions
\begin{equation}
    \nabla (\nabla  \cdot {{\bf{H}}}) - {\nabla ^2}{{\bf{H}}} + \varepsilon {\mu _0}\frac{{{\partial ^2}}}{{\partial {t^2}}}({{\bf{H}}} + {{\bf{M}}}) = 0.
    \label{eq3}
\end{equation}
For the non-magnetic regions II and III, the magnetization vanishes (i.e. $\mathbf{M}=0$). The periodically grooved metallic surface in Region III supports TM spoof surface plasmons propagating along the $x-$direction as shown in Fig. \ref{fig:1}. Following the effective-medium approach \cite{pendry2004mimicking}, the corrugated metal could be modeled as a metallic surface covered by a homogeneous anisotropic medium of thickness $d$, whose effective permittivity and permeability tensors are given by \cite{garcia-vidal2005surfaces} 
\begin{subequations}
\begin{align}
\overline \varepsilon   = \left( {\begin{array}{*{20}{c}}
{{\rm{p/}}a}&0&0\\
0&\infty &0\\
0&0&\infty 
\end{array}} \right), \\
\overline \mu   = \left( {\begin{array}{*{20}{c}}
1&0&0\\
0&{a/p}&0\\
0&0&{a/p}
\end{array}} \right).
\end{align}
\label{eq1}
\end{subequations}

For the AFM material in Region I, its magnetic response is fundamentally different from those of the nonmagnetic regions. 
For simplicity, we consider a uniaxial two-sublattice AFM insulator with its easy-axis along the $x-$axis as shown  in Fig. \ref{fig:1}. The classical ground state of the AFM layer is a Néel state with the magnetic moments on the two sublattices aligned antiparallel to each other along the $\pm x-$directions, i.e., ${{\bf{M}}_1} = {M_s}{{\bf{m}}_1}$, ${{\bf{M}}_2} = {M_s}{{\bf{m}}_2}$, where $\mathbf{m}_1$ and $\mathbf{m}_2$ are unit vectors specifying the orientations of the magnetic moments. Under a time-varying external magnetic field, the magnetic moments are driven away from their original equilibrium orientations and perform oscillations, and their dynamics are governed by the two coupled LLG equations
\begin{subequations}
    \begin{gather}
        \frac{{\partial {{\bf{m}}_1}}}{{\partial t}} =  - \gamma {{\bf{m}}_1} \times {{\bf{H}}_{1,\mathrm{eff}}} + \alpha {{\bf{m}}_1} \times \frac{{\partial {{\bf{m}}_1}}}{{\partial t}}, \\
        \frac{{\partial {{\bf{m}}_2}}}{{\partial t}} =  - \gamma {{\bf{m}}_2} \times {{\bf{H}}_{2,\mathrm{eff}}} + \alpha {{\bf{m}}_2} \times \frac{{\partial {{\bf{m}}_2}}}{{\partial t}},
    \end{gather}
    \label{eq4}
\end{subequations}
\hspace*{-0.3em}where $\gamma$ is the gyromagnetic ratio, $\alpha$ is the  Gilbert damping coefficient, and \textbf{H}$_{1,\mathrm{eff}}$, \textbf{H}$_{2,\mathrm{eff}}$ denote the effective fields acting on the two sublattices, respectively. The effective fields include the exchange field, anisotropy field, and external magnetic field, i.e. ${{\bf{H}}_{1,\mathrm{eff}}} =  - {H_{ex}}{{\bf{m}}_2} + {H_a}{m_{1x}}{{\bf{e}}_x} + h(t){{\bf{e}}_y}$, ${{\bf{H}}_{2,\mathrm{eff}}} =  - {H_{ex}}{{\bf{m}}_1} + {H_a}{m_{2x}}{{\bf{e}}_x} + h(t){{\bf{e}}_y}$.

For small-amplitude oscillations, the direction vector of the magnetic moments can be expanded into an equilibrium term and a first-order perturbation, i.e. ${{\bf{m}}_1} \approx{{\bf{e}}_x} + {m_{{1x}}}{{\bf{e}}_x} + {m_{{1y}}}{{\bf{e}}_y} + {m_{1z}}{{\bf{e}}_z}$, 
${{\bf{m}}_2} \approx  - {{\bf{e}}_x} + {m_{{2x}}}{{\bf{e}}_x} + {m_{{2y}}}{{\bf{e}}_y} + {m_{2z}}{{\bf{e}}_z}$. Numerical simulations and symmetry analysis reveal that, when the oscillations of magnetic moments on the two sublattices reach a steady state, they satisfy ${m_{1x}} + {m_{2x}} \approx 0$, ${m_{1z}} + {m_{2z}} \approx 0$ and ${m_{1y}} = {m_{2y}}$ \cite{yuan2025strong}. Thus, by defining the total magnetization as ${{\bf{M}}} = {M_s} (\bf{m}_1 + \bf{m}_2)$ and substituting the effective fields into Eq. \eqref{eq4}, the magnetization can be readily solved as
\begin{equation}
 M_x = M_y=0, {M_y}(t) = \chi_y h(t),\\
    \label{eq5}
\end{equation}
where the magnetic susceptibility reads 
\begin{equation}
    {\chi _y} \equiv \frac{{2{\gamma ^2}{H_a}{M_s}}}{{{\Omega ^2} - 2i\alpha \gamma ({H_{ex}} + {H_a})\omega  - {\omega ^2}}},
    \label{eq6}
\end{equation}
with ${\Omega } = {\gamma }\sqrt{{H_{\rm{a}}}(2{H_{ex}} + {H_a})}$ being the resonance frequency of the AFM magnons. Equation \eqref{eq5} shows that, under the influence of $h(t)e_y$, the AFM acquires a small periodically oscillating magnetic moment in the $y-$direction.

We are now ready to construct the complete form of the electromagnetic field profiles in the three regions of the hybrid structure and then derive the dispersion relation by imposing the boundary conditions at the interfaces. Considering the bound TM modes propagating along the $x-$direction shown in Fig. \ref{fig:1}, the magnetic field in each region takes the form of
\begin{subequations}
\begin{gather}
\mathbf{H}_1 = H_1 e^{ik_{1x}x - k_{1z}z} \mathbf{e}_y, \\
\mathbf{H}_2 = ( H_2^+ e^{ik_{2x}x + k_{2z}z}  + H_2^- e^{ik_{2x}x - k_{2z}z}) \mathbf{e}_y, \\
\mathbf{H}_3 = ( H_3^+ e^{ik_{3x}x + k_{3z}z}  + H_3^- e^{ik_{3x}x - k_{3z}z}) \mathbf{e}_y,
\end{gather}
\label{eq7}
\end{subequations}
\hspace*{-0.3em}where $k_{ix}$ and $k_{iz}$ are, respectively, the wavevectors along the $x$ and $z$ directions in the $i-$th region. The magnetic field in Eq.(\ref{eq7}) satisfies the wave equation described by Eq.(\ref{eq3}), thus we obtain
\begin{subequations}
    \begin{gather}
        {k_{1z}} = \sqrt {k_{1x}^2 - k_1^2(1 + \chi_y )}, \\
        {k_{2z}} = \sqrt {k_{2x}^2 - k_2^2}, 
    \end{gather}
    \label{eq8}
\end{subequations}
\hspace*{-0.5em}with $k_i = \sqrt{\epsilon_i \mu_i} \omega /c $ being the wavevector in the $i-$th region and $c$ being the speed of light. For region 3, by substituting the magnetic field expression $\mathbf{H}_3$ from Eq. \eqref{eq7} into Eq. \eqref{eq3} and imposing the effective permittivity and permeability given by Eq. \eqref{eq1}, we can obtain $k_{3z}=k_3$. The electric field in all three regions can be derived from Eq. (\ref{eq2}).  

At the interfaces $z=0$ and $z=h$, the tangential components of the electric and magnetic fields should be continuous due to the absence of surface charges and currents. At the metallic interface $z=-d$, the tangential components of electric fields vanish under the perfect-conductor approximation. As a result, the boundary conditions that the electromagnetic fields must satisfy at the interfaces can be expressed as 
\begin{subequations}
    \begin{gather}
        ({H_{1y}} - {H_{2y}}){|_{z = h}} = 0, ({E_{1x}} - {E_{2x}}){|_{z = h}} = 0, \\
        ({H_{2y}} - {H_{3y}}){|_{z = 0}} = 0, ({E_{2x}} - {E_{3x}}){|_{z = 0}} = 0, \\
        {E_{3x}}{|_{z =  - d}} = 0.
    \end{gather}
    \label{eq9}
\end{subequations}    
\hspace*{-0.3em}One can reformulate the boundary conditions as a set of linear equations
\begin{equation}
    \bf {A} \textbf{·}{\bf{ X}} = {\bf{0}},
    \label{eq10}
\end{equation}
where $\mathbf{X}= (H_{1}, H_2^+,H_2^-,H_3^+,H_3^-)^T$ and
\begin{equation}
\setlength\arraycolsep{1pt}     
\renewcommand{\arraystretch}{1.4}
\bf {A}  = \left[ {\begin{array}{*{20}{c}}
{{e^{ - {k_{1z}}h}}}&{ - {e^{{k_{2z}}h}}}&{ - {e^{ - {k_{2z}}h}}}&0&0\\
{ - \frac{{{k_{1z}}}}{{{\varepsilon _1}}}{e^{ - {k_{1z}}h}}}&{ - \frac{{{k_{2z}}}}{{{\varepsilon _2}}}{e^{{k_{2z}}h}}}&{\frac{{{k_{2z}}}}{{{\varepsilon _2}}}{e^{ - {k_{2z}}h}}}&0&0\\
0&1&1&{ - 1}&{ - 1}\\
0&{ - i\frac{{{k_{2z}}}}{{{\varepsilon _2}}}}&{i\frac{{{k_{2z}}}}{{{\varepsilon _2}}}}&{ - \frac{a}{p}{k_3}}&{\frac{a}{p}{k_3}}\\
0&0&0&{{e^{ - i{k_3}d}}}&{ - {e^{i{k_3}d}}}
\end{array}} \right].
\label{eq11}
\end{equation}
To have a non-trivial solution of electromagnetic waves, we have $det(A)=0$ and can thus derive the dispersion relation of the hybrid excitations as
\begin{equation}
   {e^{2{k_{2z}}h}} = \frac{{({k_{2z}}/{\varepsilon _2} - {k_{1z}}/{\varepsilon _1})[{k_{2z}}/{\varepsilon _2} + \frac{a}{p}{k_3}\tan ({k_3}d)]}}{{({k_{2z}}/{\varepsilon _2} + {k_{1z}}/{\varepsilon _1})[{k_{2z}}/{\varepsilon _2} - \frac{a}{p}{k_3}\tan ({k_3}d)]}}.
    \label{eq12}
\end{equation}

Equation \eqref{eq12} is usually difficult to solve analytically and must be treated numerically to obtain the hybrid-mode dispersion. Nevertheless, before performing the numerical calculations, it is instructive to derive an approximate analytical expression that explicitly reveals the underlying magnon-plasmon coupling mechanism. To this end, we interpret the hybrid structure shown in Fig. \ref{fig:1} as an effective Fabry–Pérot cavity, in which the magnetic medium and the grooved metal in the hybrid structure serve as two partially reflecting mirrors. Under this interpretation, Eq. \eqref{eq12} can be transformed into the resonance equation
\begin{equation}
    F(k_x, \omega)=1 - {r_{21}}{r_{23}}{e^{ - 2{k_{2z}}d}} = 0.
    \label{eq13}
\end{equation}
where ${r_{21}} = ({k_{2z}}/{\varepsilon _2} - {k_{1z}}/{\varepsilon _1})/({k_{2z}}/{\varepsilon _2} + {k_{1z}}/{\varepsilon _1})$ and ${r_{23}} = [{k_{2z}}/{\varepsilon _2} + \frac{a}{p}{k_3}\tan ({k_3}d)]/[{k_{2z}}/{\varepsilon _2} - \frac{a}{p}{k_3}\tan ({k_3}d)]$, are the effective reflection coefficients of the electromagnetic wave at the dielectric$|$magnet and dielectric$|$grooved-metal interfaces, respectively. The resonance equation \eqref{eq13} states that the reflections at the two interfaces are exactly compensated for by the field attenuation accumulated during a round trip inside the spacer, giving rise to a self-sustained cavity mode. 

To illustrate the coupling mechanism, we first consider a normal dielectric medium III without magnetic properties ($\chi_y=0$). The resonance condition \eqref{eq13} then reduces to $F_0(k_x,\omega_{p})=0$, which yields the dispersion relation of the bare spoof plasmon as
\begin{equation}
   \frac{{\sqrt {{k_x^2}/k_d^2 - {\omega ^2}/\omega _d^2} }}{{\omega /{\omega _d}}} = \frac{a}{p}\tan (2\pi \omega /{\omega _d}),
\end{equation}
where $k_d=2\pi /d$, $\omega_d=ck_d=2\pi c/d$. Figure \ref{fig:2}(a) shows the dispersions of the bare spoof surface plasmon for different geometric parameters of the groove. As the wave vector increases, the dispersion gradually deviates from the light line and approaches an asymptotic frequency, a characteristic feature of spoof surface plasmons. The groove geometry provides an efficient means of engineering the plasmon dispersion. On one hand, increasing the groove depth $d$ lowers the asymptotic frequency $\omega_d$ and then shifts the operating frequency toward the THz regime. On the other hand,
increasing the groove width-to-period ratio $a/p$ results in a larger deviation from the light line, indicating stronger electromagnetic confinement. Therefore, both the groove depth and the groove filling ratio determine the confinement and dispersion of the spoof surface plasmons. Their geometric tunability provides a convenient knob to match the plasmon frequency with the AFM magnon resonance.

When the magnetic layer appears, we can first expand the function $F_0(k_x,\omega)$ near the plasmon frequency $\omega_{p}$ as
\begin{equation}
    {F_0}(k_x,\omega ) = \left( {\omega  - {\omega _{p}}} \right)\frac{{\partial F_0}}{{\partial \omega }}{\Big |_{\omega  = {\omega _{p}}}}.
    \label{eq14}
\end{equation}
The magnetic response of the AFM film modifies the response condition through the susceptibility $\chi_y$. Assuming a weak magnetic perturbation, we can expand $F(k_x,\omega)$ as
\begin{equation}
    F(k_x,\omega)=F_0(k_x,\omega)+\frac{{\partial F}}{{\partial \chi_y }}{\Big |_{\chi_y  = 0}}\chi_y.
    \label{eq15}
\end{equation}

By substituting Eq. (\ref{eq14}) into Eq. (\ref{eq15}), and assuming $\alpha=0$, we derive a simple expression for the hybrid mode frequency as
\begin{equation}
    (\omega  - {\omega _{p}})(\omega  - \Omega ) =   \frac{{{\gamma ^2}{H_{\rm{a}}}{M_s}}}{\Omega }\frac{{{\partial _{\chi_y} }F{|_{\chi_y  = 0}}}}{{{\partial _\omega }{F_0}{|_{\omega  = {\omega _{p}}}}}} \equiv {g^2}.
    \label{eq17}
\end{equation}
Equation \eqref{eq17} is precisely the characteristic equation of two coherently-coupled harmonic modes. It establishes a clear connection between the phenomenological coupling strength and the electromagnetic boundary-value problems and thus gives a microscopic interpretation of the plasmon-magnon interaction. The corresponding hybrid-mode frequencies are therefore
\begin{equation}
    \omega  = \frac{{{\omega _{p}} + \Omega }}{2} + \sqrt {{{\left( {\frac{{{\omega _{p}} - \Omega }}{2}} \right)}^2} + {g^2}}.
    \label{eq18}
\end{equation}
Near the frequency crossing point of the two modes ($\Omega$), the hybrid dispersion takes on a clear avoided crossing behavior. Nevertheless, one has to rely on the numerical method to determine the coupling strength $g$ as we shall see below.

\section{Coupling spectrum and cooperativity}
\subsection{Coupling spectrum}
Now we investigate the full dispersion relation of hybrid magnon-plasmon excitations by numerically solving Eq. \eqref{eq12}. As examples, we consider the widely used AFM materials (FeF$_{2}$, MnF$_{2}$ and NiO), which span a broad frequency range from sub-THz to THz regimes. Their magnetic parameters and dielectric constants are summarized in Table \ref{table1}. 

\begin{table}[htbp]
\centering
\caption{Magnetic parameters and dielectric constants for three AFM insulators FeF$_2$, MnF$_2$, and NiO.}
\setlength{\doublerulesep}{1.5pt} 
\begin{tabular*}{\columnwidth}{@{\extracolsep{\fill}}lccc}
\hline\hline
      & FeF$_2$ & MnF$_2$ & NiO \\
\hline
H$_{ex}$ (T)   & 79 & 67 & 524 \\
H$_a$ (T) & 20  & 0.787  & 1.47 \\
M$_s$ (T) & 0.056 & 0.060 & 0.32 \\
$\Omega$ (THz)         & 10.02 & 1.64 & 6.91 \\
$\varepsilon$ & 5.4 & 6.2 & 11.9 \\
ref. & \cite{luthi1983surface,sanders1981far} & \cite{luthi1983surface,schleck2010elastic} & \cite{moriyama2019intrinsic,rao1965dielectric} \\
\hline\hline
\end{tabular*}
\label{table1}
\end{table}

\begin{figure}
    \centering
    \includegraphics[width=1\linewidth]{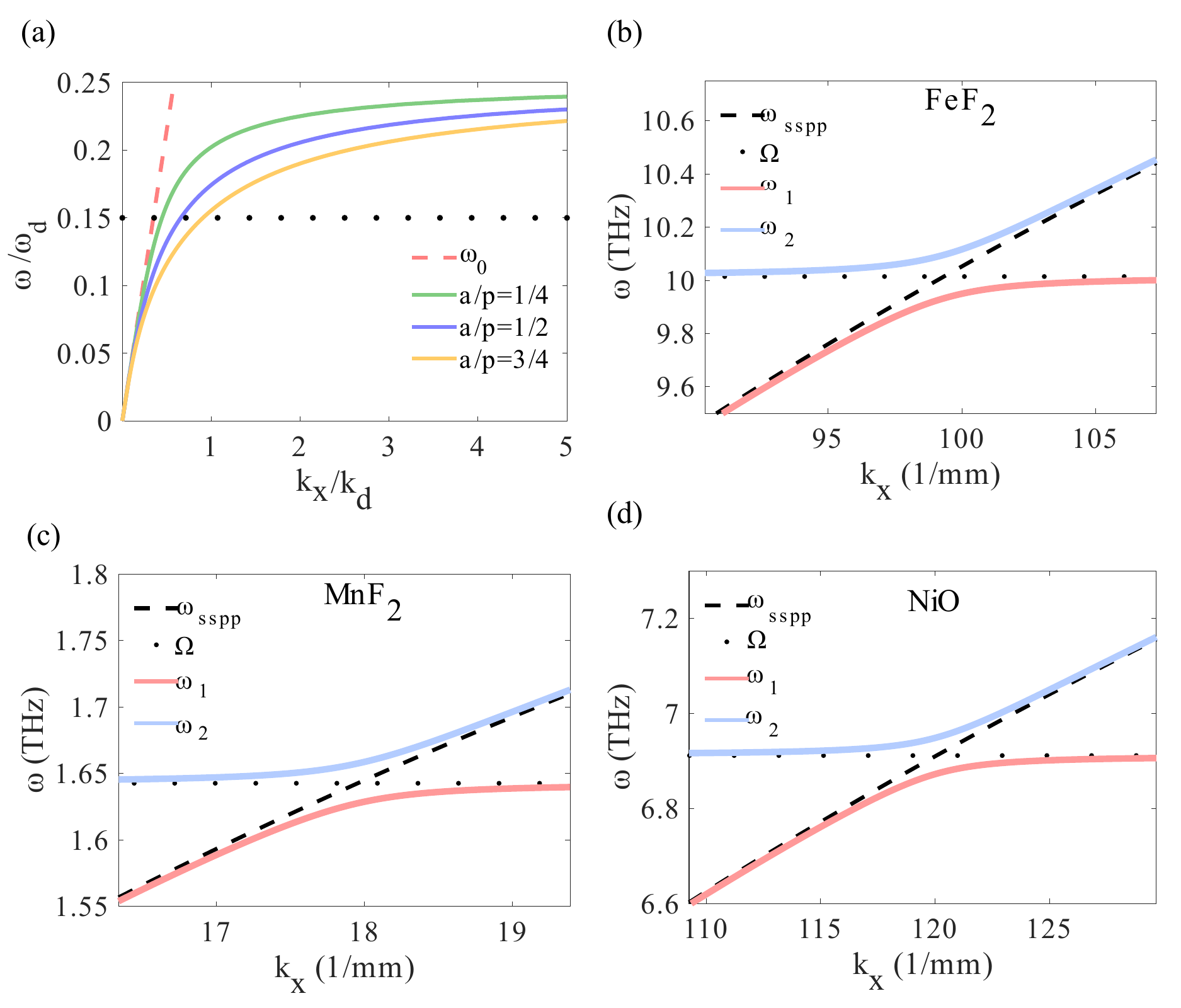}
    \caption{(a) Dispersions of bare spoof surface plasmons for different geometric parameters. (b-d) Dispersions of the hybrid magnon-plasmon excitations in the heterostructures of AFM, dielectric spacer and grooved metals. The geometric parameters are $a/p = 1/4, h=1~\mu \mathrm{m}$, and the groove depth $d$ is chosen to satisfy $\Omega/\omega_d=0.15$, tuning the spoof surface plasmon into resonance with the AFM magnons. (b) FeF$_{2}$, $d=28.2~\mu \mathrm{m}$; (c) MnF$_{2}$, $d=172.1~\mu \mathrm{m}$; (d) NiO, $d=40.9~\mu \mathrm{m}$. The magnetic parameters are shown in Table \ref{table1}.}
    \label{fig:2}
\end{figure}

Figures \ref{fig:2}(b-d) show the dispersion relations of hybrid magnon-plasmon excitations for FeF$_{2}$, MnF$_{2}$ and NiO, respectively. Here, the black dashed and dotted lines represent the uncoupled dispersion curves of spoof surface plasmons and AFM magnons, respectively, while the blue and red solid lines denote the hybrid eigenmodes of the coupled system. In all three materials, pronounced avoided crossings are observed when the two modes become resonant, suggesting a coherent coupling between magnons and surface plasmons. The resulting mode splitting ranges from several tens to more than one hundred GHz, depending on the material parameters.

\begin{figure}
    \centering
    \includegraphics[width=1\linewidth]{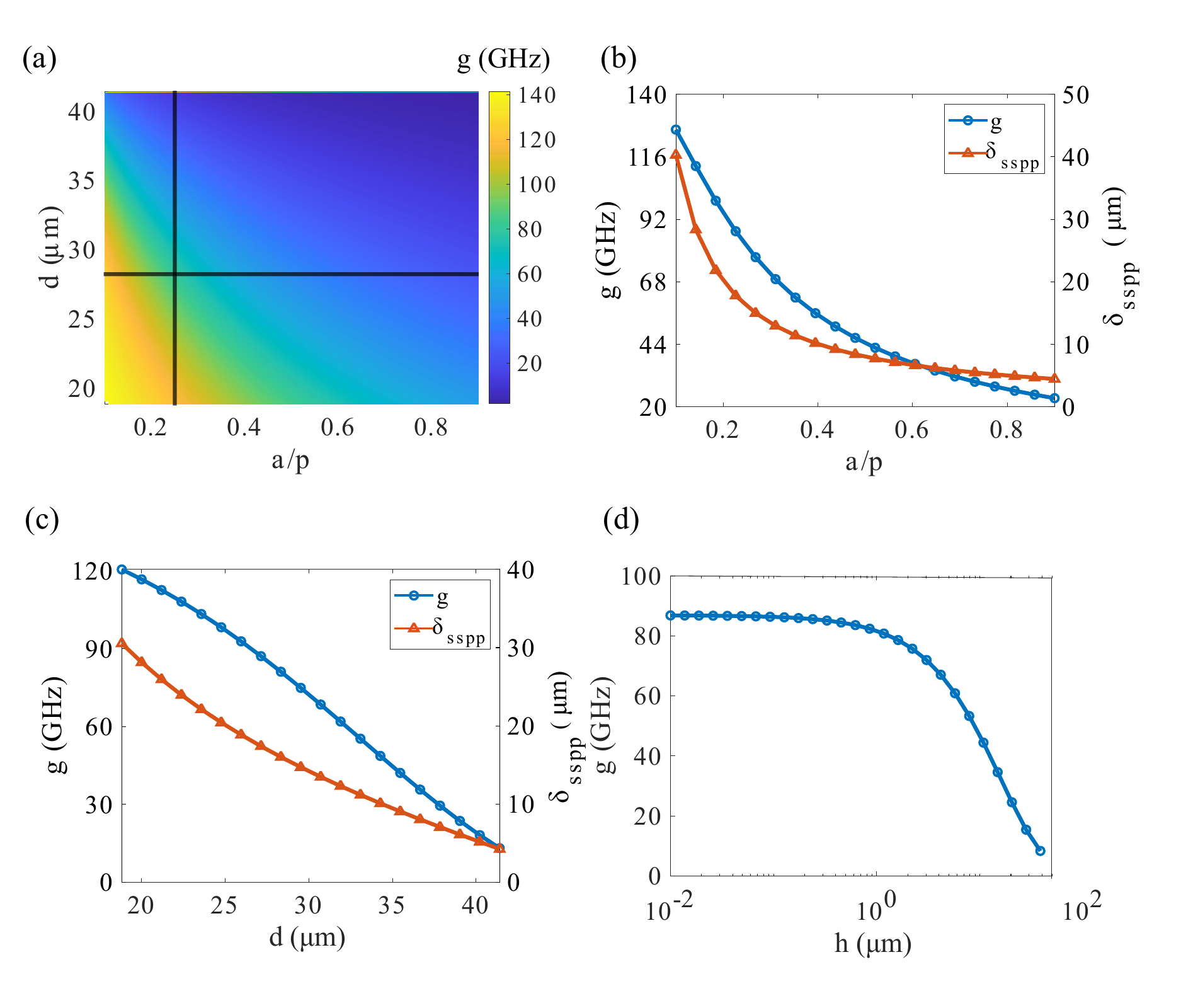}
    \caption{(a) Coupling strength as a function of the groove width-to-period ratio $a/p$ and groove depth $d$. (b) Coupling strength and penetration depth of spoof surface plasmon versus $a/p$ along the horizontal dashed line in (a) with $d=28.2~\mu$m, $h=1~\mu$m. (c) Coupling strength and penetration depth of spoof surface plasmon versus $d$ along the vertical dashed line in (a) with $a/p=1/4$, $h=1~\mu$m. (d) Coupling strength as a function of $h$ with $d=28.2~\mu$m and $a/p=1/4$.} 
    \label{fig:3}
\end{figure}

Furthermore, the coupling strength can be tailored through the geometric parameters of the hybrid structure. Taking $\mathrm{FeF}_2$ as an example, Fig. \ref{fig:3}(a) illustrates the coupling strength as a function of the groove geometry, while Fig. \ref{fig:3}(b) and \ref{fig:3}(c) show the representative dependences on ($a/p$) and ($d$), respectively. In general, decreasing either $a/p$ or $h$ enhances the magnon-plasmon coupling. Physically, this is because smaller $a/p$ and $d$ usually imply less confinement of surface plasmons and larger evanescent-field penetration into the dielectric spacer, increasing the spatial overlap between the spoof surface plasmons and AFM magnons (red lines in  Figs. \ref{fig:3}(b), \ref{fig:3}(c)). Consequently, the coupling strength increases.

Figure \ref{fig:3}(d) shows the coupling strength decreases monotonically with increasing the dielectric thickness $h$. Since both surface plasmons and the AFM magnons are localized near the interface, their coupling is mediated by the overlap of their  evanescent fields inside the dielectric spacer. Increasing the dielectric thickness reduces this mode overlap and therefore weakens the coherent coupling.  For the AFM material studies here, the magnon-plasmon coupling strength can exceed 100 GHz by adjusting the geometric parameters, which indicates a substantially efficient energy exchange between the two modes. In the following section, we further examine whether this coupling falls into the strong-coupling regime by evaluating the cooperativity.

\subsection{Cooperativity}
To assess whether magnons and plasmons reaches the strong coupling regime, we here evaluate the cooperativity defined as
\begin{equation}
    C = \frac{{4{g^2}}}{{\kappa_p \kappa_m }},
    \label{eq19}
\end{equation}
where $g$ is the coupling strength which can be directly extracted from the dispersion curves, $\kappa_m$ is the dissipation rate of magnons, characterizing the decay rate of AFM magnons energy caused by magnetic Gilbert damping. $\kappa_p$ is the decay rate of the spoof surface plasmons. The cooperativity characterizes the competition between coherent energy exchange rate and dissipation loss of the hybrid system: when $C>1$, the energy exchange rate dominates the dissipation rate, and the system operates in the strong-coupling regime; when $C<1$, the system is in the weak-coupling regime.

For a realistic metal in the terahertz regime, the electromagnetic wave penetrates into the metal over a finite skin depth. Consequently, the dissipation rate of the spoof surface plasmons arises mainly from the Ohmic losses induced by this field penetration into the metal. When calculating the spoof surface plasmon, Rusina et al. \cite{rusina2010theory} accounted for the effect of the finite skin depth of plasmons in the THz regime and obtained a relationship among the imaginary part of the surface plasmon wave vector, the skin depth and the structural parameters of the metallic microstructure as
\begin{equation}
    {\mathop{\rm Im}\nolimits} (k) \approx \frac{{{k_0}\varepsilon_d^{3/2}}}{{2\varepsilon{_g}}} \cdot \zeta \left( {\sqrt {{\varepsilon_g}} {k_0}d} \right) \cdot {\left( {\frac{a}{p}} \right)^2} \cdot \frac{{{l_s}}}{a}
    \label{eq20}
\end{equation}
where ${\varepsilon _d}$ and ${\varepsilon _{\rm{g}}}$ are the relative permittivities of the dielectric and the groove-filling medium, respectively, $\varepsilon_{d}$ is taken from Table \ref{table1} and $\varepsilon_g$ is chosen to be 1 here, ${l_s}$ is the skin depth of the metal. For copper, its skin depth in the THz frequency range is approximately 65 nm \cite{ordal1983optical}. The geometric parameters take the same values as those in Fig. \ref{fig:2}.
\begin{table}[htbp]
\centering
\caption{Magnon-plasmon coupling strength, dissipation rate and cooperativity for three AFM insulators FeF$_2$, MnF$_2$, and NiO.}
\setlength{\doublerulesep}{1.5pt} 
\begin{tabular*}{\columnwidth}{@{\extracolsep{\fill}}lccc}
\hline\hline
      & FeF$_2$ & MnF$_2$ & NiO \\
\hline
g (GHz)   & 79.23 & 14.97 & 38.12 \\
$\kappa_m$ (GHz) & 9.09 \cite{dumelow1997continuum}  & 0.99 \cite{kotthaus1972antiferromagnetic}  & 3.46 \cite{moriyama2019intrinsic} \\
$\kappa_p$ (GHz) & 27.08 & 4.78 & 27.67 \\
$C$         & 25.50 & 47.36 & 15.18 \\
\hline\hline
\end{tabular*}
\label{table2}
\end{table}

We have calculated the cooperativity for FeF$_2$, MnF$_2$, and NiO and summarize the results in Table \ref{table2}. Among the three materials, the calculated cooperativity ranges from approximately 15 to 47, confirming that the coupling between AFM magnons and spoof surface plasmons falls into the strong-coupling regime. The relative permittivity of the dielectric layer ($\epsilon_d$) strongly affects the dissipation rate of spoof plasmons. According to Eq. \eqref{eq20}, $\epsilon_d$ amplifies the imaginary part of the plasmonic wave vector through a power-law dependence of $\epsilon_d^{2/3}$, thereby increasing the dissipation rate. To reduce the losses, we can choose a medium with low permittivity. Furthermore, the cooperativity can be further enhanced by reducing the metallic skin depth or optimizing the groove geometry. Note that the groove width has to be well below the free-space wavelength to validate the effective-medium approach of spoof surface plasmons. 

\section{Discussions and conclusions.}
In conclusion, we have demonstrated strong coupling between AFM magnons and spoof surface plasmons in a planar AFM$|$dielectric$|$grooved-metal structure. The hybrid-mode dispersion exhibits a pronounced avoided crossing with a mode splitting exceeding 100 GHz, and the corresponding cooperativity values are significantly larger than unity, confirming that the hybrid systems operate in the strong‑coupling regime. Owing to the geometric tunability of spoof surface plasmons, both the resonance frequency and the coupling strength can be efficiently engineered through the groove geometry and dielectric spacer thickness. Thus, our proposal provides a flexible and experimentally accessible platform for tailoring magnon-plasmon coupling in sub-THz and THz regimes. Furthermore, our proposal may stimulate further studies to explore quantum magnonics phenomenon based on AFM systems \cite{YuanReview2022} and find promising applications in the development of hybrid magneto‑optical devices for low‑power information processing.

{\it Acknowledgments.}---This work is supported by the National Key R$\&$D Program of China (2022YFA1402700) and the National Natural Science Foundation of China (NSFC) (Grant No. 12574132).

\bibliography{magnon_spoofplasmon}

\begin{thebibliography}{41}%
\makeatletter
\providecommand \@ifxundefined [1]{%
 \@ifx{#1\undefined}
}%
\providecommand \@ifnum [1]{%
 \ifnum #1\expandafter \@firstoftwo
 \else \expandafter \@secondoftwo
 \fi
}%
\providecommand \@ifx [1]{%
 \ifx #1\expandafter \@firstoftwo
 \else \expandafter \@secondoftwo
 \fi
}%
\providecommand \natexlab [1]{#1}%
\providecommand \enquote  [1]{``#1''}%
\providecommand \bibnamefont  [1]{#1}%
\providecommand \bibfnamefont [1]{#1}%
\providecommand \citenamefont [1]{#1}%
\providecommand \href@noop [0]{\@secondoftwo}%
\providecommand \href [0]{\begingroup \@sanitize@url \@href}%
\providecommand \@href[1]{\@@startlink{#1}\@@href}%
\providecommand \@@href[1]{\endgroup#1\@@endlink}%
\providecommand \@sanitize@url [0]{\catcode `\\12\catcode `\$12\catcode `\&12\catcode `\#12\catcode `\^12\catcode `\_12\catcode `\%12\relax}%
\providecommand \@@startlink[1]{}%
\providecommand \@@endlink[0]{}%
\providecommand \url  [0]{\begingroup\@sanitize@url \@url }%
\providecommand \@url [1]{\endgroup\@href {#1}{\urlprefix }}%
\providecommand \urlprefix  [0]{URL }%
\providecommand \Eprint [0]{\href }%
\providecommand \doibase [0]{https://doi.org/}%
\providecommand \selectlanguage [0]{\@gobble}%
\providecommand \bibinfo  [0]{\@secondoftwo}%
\providecommand \bibfield  [0]{\@secondoftwo}%
\providecommand \translation [1]{[#1]}%
\providecommand \BibitemOpen [0]{}%
\providecommand \bibitemStop [0]{}%
\providecommand \bibitemNoStop [0]{.\EOS\space}%
\providecommand \EOS [0]{\spacefactor3000\relax}%
\providecommand \BibitemShut  [1]{\csname bibitem#1\endcsname}%
\let\auto@bib@innerbib\@empty
\bibitem [{\citenamefont {Yuan}\ \emph {et~al.}(2022)\citenamefont {Yuan}, \citenamefont {Cao}, \citenamefont {Kamra}, \citenamefont {Duine},\ and\ \citenamefont {Yan}}]{YuanReview2022}%
  \BibitemOpen
  \bibfield  {author} {\bibinfo {author} {\bibfnamefont {H.}~\bibnamefont {Yuan}}, \bibinfo {author} {\bibfnamefont {Y.}~\bibnamefont {Cao}}, \bibinfo {author} {\bibfnamefont {A.}~\bibnamefont {Kamra}}, \bibinfo {author} {\bibfnamefont {R.~A.}\ \bibnamefont {Duine}},\ and\ \bibinfo {author} {\bibfnamefont {P.}~\bibnamefont {Yan}},\ }\bibfield  {title} {\bibinfo {title} {Quantum magnonics: When magnon spintronics meets quantum information science},\ }\href {https://doi.org/https://doi.org/10.1016/j.physrep.2022.03.002} {\bibfield  {journal} {\bibinfo  {journal} {Physics Reports}\ }\textbf {\bibinfo {volume} {965}},\ \bibinfo {pages} {1} (\bibinfo {year} {2022})}\BibitemShut {NoStop}%
\bibitem [{\citenamefont {{Zare Rameshti}}\ \emph {et~al.}(2022)\citenamefont {{Zare Rameshti}}, \citenamefont {{Viola Kusminskiy}}, \citenamefont {Haigh}, \citenamefont {Usami}, \citenamefont {Lachance-Quirion}, \citenamefont {Nakamura}, \citenamefont {Hu}, \citenamefont {Tang}, \citenamefont {Bauer},\ and\ \citenamefont {Blanter}}]{BabakReview20221}%
  \BibitemOpen
  \bibfield  {author} {\bibinfo {author} {\bibfnamefont {B.}~\bibnamefont {{Zare Rameshti}}}, \bibinfo {author} {\bibfnamefont {S.}~\bibnamefont {{Viola Kusminskiy}}}, \bibinfo {author} {\bibfnamefont {J.~A.}\ \bibnamefont {Haigh}}, \bibinfo {author} {\bibfnamefont {K.}~\bibnamefont {Usami}}, \bibinfo {author} {\bibfnamefont {D.}~\bibnamefont {Lachance-Quirion}}, \bibinfo {author} {\bibfnamefont {Y.}~\bibnamefont {Nakamura}}, \bibinfo {author} {\bibfnamefont {C.-M.}\ \bibnamefont {Hu}}, \bibinfo {author} {\bibfnamefont {H.~X.}\ \bibnamefont {Tang}}, \bibinfo {author} {\bibfnamefont {G.~E.}\ \bibnamefont {Bauer}},\ and\ \bibinfo {author} {\bibfnamefont {Y.~M.}\ \bibnamefont {Blanter}},\ }\bibfield  {title} {\bibinfo {title} {Cavity magnonics},\ }\href {https://doi.org/https://doi.org/10.1016/j.physrep.2022.06.001} {\bibfield  {journal} {\bibinfo  {journal} {Physics Reports}\ }\textbf {\bibinfo {volume} {979}},\ \bibinfo {pages} {1} (\bibinfo {year} {2022})},\ \bibinfo {note} {cavity Magnonics}\BibitemShut
  {NoStop}%
\bibitem [{\citenamefont {Li}\ \emph {et~al.}(2020)\citenamefont {Li}, \citenamefont {Zhang}, \citenamefont {Tyberkevych}, \citenamefont {Kwok}, \citenamefont {Hoffmann},\ and\ \citenamefont {Novosad}}]{LiReview2020}%
  \BibitemOpen
  \bibfield  {author} {\bibinfo {author} {\bibfnamefont {Y.}~\bibnamefont {Li}}, \bibinfo {author} {\bibfnamefont {W.}~\bibnamefont {Zhang}}, \bibinfo {author} {\bibfnamefont {V.}~\bibnamefont {Tyberkevych}}, \bibinfo {author} {\bibfnamefont {W.-K.}\ \bibnamefont {Kwok}}, \bibinfo {author} {\bibfnamefont {A.}~\bibnamefont {Hoffmann}},\ and\ \bibinfo {author} {\bibfnamefont {V.}~\bibnamefont {Novosad}},\ }\bibfield  {title} {\bibinfo {title} {Hybrid magnonics: Physics, circuits, and applications for coherent information processing},\ }\href {https://doi.org/10.1063/5.0020277} {\bibfield  {journal} {\bibinfo  {journal} {Journal of Applied Physics}\ }\textbf {\bibinfo {volume} {128}},\ \bibinfo {pages} {130902} (\bibinfo {year} {2020})}\BibitemShut {NoStop}%
\bibitem [{\citenamefont {Zuo}\ \emph {et~al.}(2024)\citenamefont {Zuo}, \citenamefont {Fan}, \citenamefont {Qian}, \citenamefont {Ding}, \citenamefont {Tan}, \citenamefont {Xiong},\ and\ \citenamefont {Li}}]{ZuoNJP2024}%
  \BibitemOpen
  \bibfield  {author} {\bibinfo {author} {\bibfnamefont {X.}~\bibnamefont {Zuo}}, \bibinfo {author} {\bibfnamefont {Z.-Y.}\ \bibnamefont {Fan}}, \bibinfo {author} {\bibfnamefont {H.}~\bibnamefont {Qian}}, \bibinfo {author} {\bibfnamefont {M.-S.}\ \bibnamefont {Ding}}, \bibinfo {author} {\bibfnamefont {H.}~\bibnamefont {Tan}}, \bibinfo {author} {\bibfnamefont {H.}~\bibnamefont {Xiong}},\ and\ \bibinfo {author} {\bibfnamefont {J.}~\bibnamefont {Li}},\ }\bibfield  {title} {\bibinfo {title} {Cavity magnomechanics: from classical to quantum},\ }\href {https://doi.org/10.1088/1367-2630/ad327c} {\bibfield  {journal} {\bibinfo  {journal} {New Journal of Physics}\ }\textbf {\bibinfo {volume} {26}},\ \bibinfo {pages} {031201} (\bibinfo {year} {2024})}\BibitemShut {NoStop}%
\bibitem [{\citenamefont {Soykal}\ and\ \citenamefont {Flatt\'e}(2010)}]{PhysRevLett.104.077202}%
  \BibitemOpen
  \bibfield  {author} {\bibinfo {author} {\bibfnamefont {O.~O.}\ \bibnamefont {Soykal}}\ and\ \bibinfo {author} {\bibfnamefont {M.~E.}\ \bibnamefont {Flatt\'e}},\ }\bibfield  {title} {\bibinfo {title} {Strong field interactions between a nanomagnet and a photonic cavity},\ }\href {https://doi.org/10.1103/PhysRevLett.104.077202} {\bibfield  {journal} {\bibinfo  {journal} {Phys. Rev. Lett.}\ }\textbf {\bibinfo {volume} {104}},\ \bibinfo {pages} {077202} (\bibinfo {year} {2010})}\BibitemShut {NoStop}%
\bibitem [{\citenamefont {Huebl}\ \emph {et~al.}(2013)\citenamefont {Huebl}, \citenamefont {Zollitsch}, \citenamefont {Lotze}, \citenamefont {Hocke}, \citenamefont {Greifenstein}, \citenamefont {Marx}, \citenamefont {Gross},\ and\ \citenamefont {Goennenwein}}]{PhysRevLett.111.127003}%
  \BibitemOpen
  \bibfield  {author} {\bibinfo {author} {\bibfnamefont {H.}~\bibnamefont {Huebl}}, \bibinfo {author} {\bibfnamefont {C.~W.}\ \bibnamefont {Zollitsch}}, \bibinfo {author} {\bibfnamefont {J.}~\bibnamefont {Lotze}}, \bibinfo {author} {\bibfnamefont {F.}~\bibnamefont {Hocke}}, \bibinfo {author} {\bibfnamefont {M.}~\bibnamefont {Greifenstein}}, \bibinfo {author} {\bibfnamefont {A.}~\bibnamefont {Marx}}, \bibinfo {author} {\bibfnamefont {R.}~\bibnamefont {Gross}},\ and\ \bibinfo {author} {\bibfnamefont {S.~T.~B.}\ \bibnamefont {Goennenwein}},\ }\bibfield  {title} {\bibinfo {title} {High cooperativity in coupled microwave resonator ferrimagnetic insulator hybrids},\ }\href {https://doi.org/10.1103/PhysRevLett.111.127003} {\bibfield  {journal} {\bibinfo  {journal} {Phys. Rev. Lett.}\ }\textbf {\bibinfo {volume} {111}},\ \bibinfo {pages} {127003} (\bibinfo {year} {2013})}\BibitemShut {NoStop}%
\bibitem [{\citenamefont {Zhang}\ \emph {et~al.}(2016)\citenamefont {Zhang}, \citenamefont {Zou}, \citenamefont {Jiang},\ and\ \citenamefont {Tang}}]{zhang2016cavity}%
  \BibitemOpen
  \bibfield  {author} {\bibinfo {author} {\bibfnamefont {X.}~\bibnamefont {Zhang}}, \bibinfo {author} {\bibfnamefont {C.-L.}\ \bibnamefont {Zou}}, \bibinfo {author} {\bibfnamefont {L.}~\bibnamefont {Jiang}},\ and\ \bibinfo {author} {\bibfnamefont {H.~X.}\ \bibnamefont {Tang}},\ }\bibfield  {title} {\bibinfo {title} {Cavity magnomechanics},\ }\bibfield  {journal} {\bibinfo  {journal} {Science Advances}\ }\textbf {\bibinfo {volume} {2}},\ \href {https://doi.org/10.1126/sciadv.1501286} {10.1126/sciadv.1501286} (\bibinfo {year} {2016})\BibitemShut {NoStop}%
\bibitem [{\citenamefont {Tabuchi}\ \emph {et~al.}(2015)\citenamefont {Tabuchi}, \citenamefont {Ishino}, \citenamefont {Noguchi}, \citenamefont {Ishikawa}, \citenamefont {Yamazaki}, \citenamefont {Usami},\ and\ \citenamefont {Nakamura}}]{tabuchi2015coherent}%
  \BibitemOpen
  \bibfield  {author} {\bibinfo {author} {\bibfnamefont {Y.}~\bibnamefont {Tabuchi}}, \bibinfo {author} {\bibfnamefont {S.}~\bibnamefont {Ishino}}, \bibinfo {author} {\bibfnamefont {A.}~\bibnamefont {Noguchi}}, \bibinfo {author} {\bibfnamefont {T.}~\bibnamefont {Ishikawa}}, \bibinfo {author} {\bibfnamefont {R.}~\bibnamefont {Yamazaki}}, \bibinfo {author} {\bibfnamefont {K.}~\bibnamefont {Usami}},\ and\ \bibinfo {author} {\bibfnamefont {Y.}~\bibnamefont {Nakamura}},\ }\bibfield  {title} {\bibinfo {title} {Coherent coupling between a ferromagnetic magnon and a superconducting qubit},\ }\href {https://doi.org/10.1126/science.aaa3693} {\bibfield  {journal} {\bibinfo  {journal} {Science}\ }\textbf {\bibinfo {volume} {349}},\ \bibinfo {pages} {405} (\bibinfo {year} {2015})}\BibitemShut {NoStop}%
\bibitem [{\citenamefont {Lachance-Quirion}\ \emph {et~al.}(2020)\citenamefont {Lachance-Quirion}, \citenamefont {Wolski}, \citenamefont {Tabuchi}, \citenamefont {Kono}, \citenamefont {Usami},\ and\ \citenamefont {Nakamura}}]{lachance2020entanglement}%
  \BibitemOpen
  \bibfield  {author} {\bibinfo {author} {\bibfnamefont {D.}~\bibnamefont {Lachance-Quirion}}, \bibinfo {author} {\bibfnamefont {S.~P.}\ \bibnamefont {Wolski}}, \bibinfo {author} {\bibfnamefont {Y.}~\bibnamefont {Tabuchi}}, \bibinfo {author} {\bibfnamefont {S.}~\bibnamefont {Kono}}, \bibinfo {author} {\bibfnamefont {K.}~\bibnamefont {Usami}},\ and\ \bibinfo {author} {\bibfnamefont {Y.}~\bibnamefont {Nakamura}},\ }\bibfield  {title} {\bibinfo {title} {Entanglement-based single-shot detection of a single magnon with a superconducting qubit},\ }\href {https://doi.org/10.1126/science.aaz9236} {\bibfield  {journal} {\bibinfo  {journal} {Science}\ }\textbf {\bibinfo {volume} {367}},\ \bibinfo {pages} {425} (\bibinfo {year} {2020})}\BibitemShut {NoStop}%
\bibitem [{\citenamefont {Xu}\ \emph {et~al.}(2023)\citenamefont {Xu}, \citenamefont {Gu}, \citenamefont {Li}, \citenamefont {Weng}, \citenamefont {Wang}, \citenamefont {Li}, \citenamefont {Wang}, \citenamefont {Zhu},\ and\ \citenamefont {You}}]{XuPRL2023}%
  \BibitemOpen
  \bibfield  {author} {\bibinfo {author} {\bibfnamefont {D.}~\bibnamefont {Xu}}, \bibinfo {author} {\bibfnamefont {X.-K.}\ \bibnamefont {Gu}}, \bibinfo {author} {\bibfnamefont {H.-K.}\ \bibnamefont {Li}}, \bibinfo {author} {\bibfnamefont {Y.-C.}\ \bibnamefont {Weng}}, \bibinfo {author} {\bibfnamefont {Y.-P.}\ \bibnamefont {Wang}}, \bibinfo {author} {\bibfnamefont {J.}~\bibnamefont {Li}}, \bibinfo {author} {\bibfnamefont {H.}~\bibnamefont {Wang}}, \bibinfo {author} {\bibfnamefont {S.-Y.}\ \bibnamefont {Zhu}},\ and\ \bibinfo {author} {\bibfnamefont {J.~Q.}\ \bibnamefont {You}},\ }\bibfield  {title} {\bibinfo {title} {Quantum control of a single magnon in a macroscopic spin system},\ }\href {https://doi.org/10.1103/PhysRevLett.130.193603} {\bibfield  {journal} {\bibinfo  {journal} {Phys. Rev. Lett.}\ }\textbf {\bibinfo {volume} {130}},\ \bibinfo {pages} {193603} (\bibinfo {year} {2023})}\BibitemShut {NoStop}%
\bibitem [{\citenamefont {Weng}\ \emph {et~al.}(2026)\citenamefont {Weng}, \citenamefont {Xu}, \citenamefont {Chen}, \citenamefont {Tan}, \citenamefont {Gu}, \citenamefont {Li}, \citenamefont {Yu}, \citenamefont {Zhu}, \citenamefont {Hu}, \citenamefont {Nori},\ and\ \citenamefont {You}}]{WengNC2026}%
  \BibitemOpen
  \bibfield  {author} {\bibinfo {author} {\bibfnamefont {Y.-C.}\ \bibnamefont {Weng}}, \bibinfo {author} {\bibfnamefont {D.}~\bibnamefont {Xu}}, \bibinfo {author} {\bibfnamefont {Z.}~\bibnamefont {Chen}}, \bibinfo {author} {\bibfnamefont {L.-Z.}\ \bibnamefont {Tan}}, \bibinfo {author} {\bibfnamefont {X.-K.}\ \bibnamefont {Gu}}, \bibinfo {author} {\bibfnamefont {J.}~\bibnamefont {Li}}, \bibinfo {author} {\bibfnamefont {H.-F.}\ \bibnamefont {Yu}}, \bibinfo {author} {\bibfnamefont {S.-Y.}\ \bibnamefont {Zhu}}, \bibinfo {author} {\bibfnamefont {X.}~\bibnamefont {Hu}}, \bibinfo {author} {\bibfnamefont {F.}~\bibnamefont {Nori}},\ and\ \bibinfo {author} {\bibfnamefont {J.~Q.}\ \bibnamefont {You}},\ }\bibfield  {title} {\bibinfo {title} {Magnon squeezing in the quantum regime},\ }\href {https://doi.org/10.1038/s41467-026-69312-4} {\bibfield  {journal} {\bibinfo  {journal} {Nature Communications}\ }\textbf {\bibinfo {volume} {17}},\ \bibinfo {pages} {2679} (\bibinfo {year} {2026})}\BibitemShut {NoStop}%
\bibitem [{\citenamefont {Bludov}\ \emph {et~al.}(2019)\citenamefont {Bludov}, \citenamefont {Gomes}, \citenamefont {Farias}, \citenamefont {Fernández-Rossier}, \citenamefont {Vasilevskiy},\ and\ \citenamefont {Peres}}]{bludov2019hybrid}%
  \BibitemOpen
  \bibfield  {author} {\bibinfo {author} {\bibfnamefont {Y.~V.}\ \bibnamefont {Bludov}}, \bibinfo {author} {\bibfnamefont {J.~N.}\ \bibnamefont {Gomes}}, \bibinfo {author} {\bibfnamefont {G.~A.}\ \bibnamefont {Farias}}, \bibinfo {author} {\bibfnamefont {J.}~\bibnamefont {Fernández-Rossier}}, \bibinfo {author} {\bibfnamefont {M.~I.}\ \bibnamefont {Vasilevskiy}},\ and\ \bibinfo {author} {\bibfnamefont {N.~M.~R.}\ \bibnamefont {Peres}},\ }\bibfield  {title} {\bibinfo {title} {Hybrid plasmon-magnon polaritons in graphene-antiferromagnet heterostructures},\ }\href {https://doi.org/10.1088/2053-1583/ab2513} {\bibfield  {journal} {\bibinfo  {journal} {2D Materials}\ }\textbf {\bibinfo {volume} {6}},\ \bibinfo {pages} {045003} (\bibinfo {year} {2019})}\BibitemShut {NoStop}%
\bibitem [{\citenamefont {Costa}\ \emph {et~al.}(2023)\citenamefont {Costa}, \citenamefont {Vasilevskiy}, \citenamefont {Fernández-Rossier},\ and\ \citenamefont {Peres}}]{costa2023strongly}%
  \BibitemOpen
  \bibfield  {author} {\bibinfo {author} {\bibfnamefont {A.~T.}\ \bibnamefont {Costa}}, \bibinfo {author} {\bibfnamefont {M.~I.}\ \bibnamefont {Vasilevskiy}}, \bibinfo {author} {\bibfnamefont {J.}~\bibnamefont {Fernández-Rossier}},\ and\ \bibinfo {author} {\bibfnamefont {N.~M.~R.}\ \bibnamefont {Peres}},\ }\bibfield  {title} {\bibinfo {title} {Strongly coupled magnon–plasmon polaritons in graphene-two-dimensional ferromagnet heterostructures},\ }\href {https://doi.org/10.1021/acs.nanolett.3c00907} {\bibfield  {journal} {\bibinfo  {journal} {Nano Letters}\ }\textbf {\bibinfo {volume} {23}},\ \bibinfo {pages} {4510} (\bibinfo {year} {2023})}\BibitemShut {NoStop}%
\bibitem [{\citenamefont {Yuan}\ \emph {et~al.}(2025)\citenamefont {Yuan}, \citenamefont {Blanter},\ and\ \citenamefont {Lin}}]{yuan2025strong}%
  \BibitemOpen
  \bibfield  {author} {\bibinfo {author} {\bibfnamefont {H.~Y.}\ \bibnamefont {Yuan}}, \bibinfo {author} {\bibfnamefont {Y.~M.}\ \bibnamefont {Blanter}},\ and\ \bibinfo {author} {\bibfnamefont {H.~Q.}\ \bibnamefont {Lin}},\ }\bibfield  {title} {\bibinfo {title} {Strong and tunable coupling between antiferromagnetic magnons and surface plasmons},\ }\bibfield  {journal} {\bibinfo  {journal} {Physical Review B}\ }\textbf {\bibinfo {volume} {111}},\ \href {https://doi.org/10.1103/physrevb.111.024422} {10.1103/physrevb.111.024422} (\bibinfo {year} {2025})\BibitemShut {NoStop}%
\bibitem [{\citenamefont {Dyrdał}\ \emph {et~al.}(2023)\citenamefont {Dyrdał}, \citenamefont {Qaiumzadeh}, \citenamefont {Brataas},\ and\ \citenamefont {Barnaś}}]{dyrda2023magnonplasmon}%
  \BibitemOpen
  \bibfield  {author} {\bibinfo {author} {\bibfnamefont {A.}~\bibnamefont {Dyrdał}}, \bibinfo {author} {\bibfnamefont {A.}~\bibnamefont {Qaiumzadeh}}, \bibinfo {author} {\bibfnamefont {A.}~\bibnamefont {Brataas}},\ and\ \bibinfo {author} {\bibfnamefont {J.}~\bibnamefont {Barnaś}},\ }\bibfield  {title} {\bibinfo {title} {Magnon-plasmon hybridization mediated by spin-orbit interaction in magnetic materials},\ }\bibfield  {journal} {\bibinfo  {journal} {Physical Review B}\ }\textbf {\bibinfo {volume} {108}},\ \href {https://doi.org/10.1103/physrevb.108.045414} {10.1103/physrevb.108.045414} (\bibinfo {year} {2023})\BibitemShut {NoStop}%
\bibitem [{\citenamefont {Xiong}\ \emph {et~al.}(2024)\citenamefont {Xiong}, \citenamefont {Christy}, \citenamefont {Yan}, \citenamefont {Pishehvar}, \citenamefont {Mahdi}, \citenamefont {Wu}, \citenamefont {Cahoon}, \citenamefont {Yang}, \citenamefont {Hamilton}, \citenamefont {Zhang},\ and\ \citenamefont {Zhang}}]{xiong2024hybrid}%
  \BibitemOpen
  \bibfield  {author} {\bibinfo {author} {\bibfnamefont {Y.}~\bibnamefont {Xiong}}, \bibinfo {author} {\bibfnamefont {A.}~\bibnamefont {Christy}}, \bibinfo {author} {\bibfnamefont {Z.}~\bibnamefont {Yan}}, \bibinfo {author} {\bibfnamefont {A.}~\bibnamefont {Pishehvar}}, \bibinfo {author} {\bibfnamefont {M.}~\bibnamefont {Mahdi}}, \bibinfo {author} {\bibfnamefont {J.}~\bibnamefont {Wu}}, \bibinfo {author} {\bibfnamefont {J.~F.}\ \bibnamefont {Cahoon}}, \bibinfo {author} {\bibfnamefont {B.}~\bibnamefont {Yang}}, \bibinfo {author} {\bibfnamefont {M.~C.}\ \bibnamefont {Hamilton}}, \bibinfo {author} {\bibfnamefont {X.}~\bibnamefont {Zhang}},\ and\ \bibinfo {author} {\bibfnamefont {W.}~\bibnamefont {Zhang}},\ }\bibfield  {title} {\bibinfo {title} {Hybrid magnonics with localized spoof surface-plasmon polaritons},\ }\bibfield  {journal} {\bibinfo  {journal} {Physical Review Applied}\ }\textbf {\bibinfo {volume} {22}},\ \href {https://doi.org/10.1103/physrevapplied.22.034009} {10.1103/physrevapplied.22.034009}
  (\bibinfo {year} {2024})\BibitemShut {NoStop}%
\bibitem [{\citenamefont {Yuan}\ and\ \citenamefont {Blanter}(2024)}]{yuan2024breaking}%
  \BibitemOpen
  \bibfield  {author} {\bibinfo {author} {\bibfnamefont {H.}~\bibnamefont {Yuan}}\ and\ \bibinfo {author} {\bibfnamefont {Y.~M.}\ \bibnamefont {Blanter}},\ }\bibfield  {title} {\bibinfo {title} {Breaking surface-plasmon excitation constraint via surface spin waves},\ }\bibfield  {journal} {\bibinfo  {journal} {Physical Review Letters}\ }\textbf {\bibinfo {volume} {133}},\ \href {https://doi.org/10.1103/physrevlett.133.156703} {10.1103/physrevlett.133.156703} (\bibinfo {year} {2024})\BibitemShut {NoStop}%
\bibitem [{\citenamefont {Liu}\ \emph {et~al.}(2024)\citenamefont {Liu}, \citenamefont {Li}, \citenamefont {Fu},\ and\ \citenamefont {Wang}}]{liu2024ghost}%
  \BibitemOpen
  \bibfield  {author} {\bibinfo {author} {\bibfnamefont {Q.}~\bibnamefont {Liu}}, \bibinfo {author} {\bibfnamefont {Y.}~\bibnamefont {Li}}, \bibinfo {author} {\bibfnamefont {S.}~\bibnamefont {Fu}},\ and\ \bibinfo {author} {\bibfnamefont {X.-Z.}\ \bibnamefont {Wang}},\ }\bibfield  {title} {\bibinfo {title} {Ghost surface magnon-plasmon polariton in antiferromagnets covered with graphene monolayer},\ }\href {https://doi.org/10.1364/oe.524684} {\bibfield  {journal} {\bibinfo  {journal} {Optics Express}\ }\textbf {\bibinfo {volume} {32}},\ \bibinfo {pages} {30687} (\bibinfo {year} {2024})}\BibitemShut {NoStop}%
\bibitem [{\citenamefont {Kuznetsov}\ \emph {et~al.}(2025)\citenamefont {Kuznetsov}, \citenamefont {Qin}, \citenamefont {Flajšman},\ and\ \citenamefont {van Dijken}}]{kuznetsov2025optical}%
  \BibitemOpen
  \bibfield  {author} {\bibinfo {author} {\bibfnamefont {N.}~\bibnamefont {Kuznetsov}}, \bibinfo {author} {\bibfnamefont {H.}~\bibnamefont {Qin}}, \bibinfo {author} {\bibfnamefont {L.}~\bibnamefont {Flajšman}},\ and\ \bibinfo {author} {\bibfnamefont {S.}~\bibnamefont {van Dijken}},\ }\bibfield  {title} {\bibinfo {title} {Optical control of spin waves in hybrid magnonic-plasmonic structures},\ }\bibfield  {journal} {\bibinfo  {journal} {Science Advances}\ }\textbf {\bibinfo {volume} {11}},\ \href {https://doi.org/10.1126/sciadv.ads2420} {10.1126/sciadv.ads2420} (\bibinfo {year} {2025})\BibitemShut {NoStop}%
\bibitem [{\citenamefont {Qian}\ \emph {et~al.}(2025)\citenamefont {Qian}, \citenamefont {Hong}, \citenamefont {Wang}, \citenamefont {Wu}, \citenamefont {Yang}, \citenamefont {Hu}, \citenamefont {You},\ and\ \citenamefont {Wang}}]{qian2025unidirectional}%
  \BibitemOpen
  \bibfield  {author} {\bibinfo {author} {\bibfnamefont {J.}~\bibnamefont {Qian}}, \bibinfo {author} {\bibfnamefont {Q.}~\bibnamefont {Hong}}, \bibinfo {author} {\bibfnamefont {Z.-Y.}\ \bibnamefont {Wang}}, \bibinfo {author} {\bibfnamefont {W.-X.}\ \bibnamefont {Wu}}, \bibinfo {author} {\bibfnamefont {Y.}~\bibnamefont {Yang}}, \bibinfo {author} {\bibfnamefont {C.-M.}\ \bibnamefont {Hu}}, \bibinfo {author} {\bibfnamefont {J.-Q.}\ \bibnamefont {You}},\ and\ \bibinfo {author} {\bibfnamefont {Y.-P.}\ \bibnamefont {Wang}},\ }\bibfield  {title} {\bibinfo {title} {Unidirectional perfect absorption induced by chiral coupling in spin-momentum locked waveguide magnonics},\ }\bibfield  {journal} {\bibinfo  {journal} {Nature Communications}\ }\textbf {\bibinfo {volume} {16}},\ \href {https://doi.org/10.1038/s41467-025-63305-5} {10.1038/s41467-025-63305-5} (\bibinfo {year} {2025})\BibitemShut {NoStop}%
\bibitem [{\citenamefont {Gunnink}\ and\ \citenamefont {Mook}(2026)}]{PieterPRB2026}%
  \BibitemOpen
  \bibfield  {author} {\bibinfo {author} {\bibfnamefont {P.~M.}\ \bibnamefont {Gunnink}}\ and\ \bibinfo {author} {\bibfnamefont {A.}~\bibnamefont {Mook}},\ }\bibfield  {title} {\bibinfo {title} {Coupling of plasmons to the two-magnon continuum in antiferromagnets},\ }\href {https://doi.org/10.1103/2mml-3bbv} {\bibfield  {journal} {\bibinfo  {journal} {Phys. Rev. B}\ }\textbf {\bibinfo {volume} {113}},\ \bibinfo {pages} {094454} (\bibinfo {year} {2026})}\BibitemShut {NoStop}%
\bibitem [{\citenamefont {Hirosawa}\ \emph {et~al.}(2026)\citenamefont {Hirosawa}, \citenamefont {Gunnink},\ and\ \citenamefont {Mook}}]{HiroPRB2026}%
  \BibitemOpen
  \bibfield  {author} {\bibinfo {author} {\bibfnamefont {T.}~\bibnamefont {Hirosawa}}, \bibinfo {author} {\bibfnamefont {P.~M.}\ \bibnamefont {Gunnink}},\ and\ \bibinfo {author} {\bibfnamefont {A.}~\bibnamefont {Mook}},\ }\bibfield  {title} {\bibinfo {title} {Topological magnon-plasmon hybrids},\ }\href {https://doi.org/10.1103/9csq-48k7} {\bibfield  {journal} {\bibinfo  {journal} {Phys. Rev. B}\ }\textbf {\bibinfo {volume} {113}},\ \bibinfo {pages} {L180404} (\bibinfo {year} {2026})}\BibitemShut {NoStop}%
\bibitem [{\citenamefont {Barnes}\ \emph {et~al.}(2003)\citenamefont {Barnes}, \citenamefont {Dereux},\ and\ \citenamefont {Ebbesen}}]{williaml.barnes2003surface}%
  \BibitemOpen
  \bibfield  {author} {\bibinfo {author} {\bibfnamefont {W.~L.}\ \bibnamefont {Barnes}}, \bibinfo {author} {\bibfnamefont {A.}~\bibnamefont {Dereux}},\ and\ \bibinfo {author} {\bibfnamefont {T.~W.}\ \bibnamefont {Ebbesen}},\ }\bibfield  {title} {\bibinfo {title} {Surface plasmon subwavelength optics},\ }\href {https://doi.org/10.1038/nature01937} {\bibfield  {journal} {\bibinfo  {journal} {Nature}\ }\textbf {\bibinfo {volume} {424}},\ \bibinfo {pages} {824} (\bibinfo {year} {2003})}\BibitemShut {NoStop}%
\bibitem [{\citenamefont {Zhang}\ \emph {et~al.}(2012)\citenamefont {Zhang}, \citenamefont {Zhang},\ and\ \citenamefont {Xu}}]{junxizhang2012surface}%
  \BibitemOpen
  \bibfield  {author} {\bibinfo {author} {\bibfnamefont {J.}~\bibnamefont {Zhang}}, \bibinfo {author} {\bibfnamefont {L.}~\bibnamefont {Zhang}},\ and\ \bibinfo {author} {\bibfnamefont {W.}~\bibnamefont {Xu}},\ }\bibfield  {title} {\bibinfo {title} {Surface plasmon polaritons: physics and applications},\ }\href {https://doi.org/10.1088/0022-3727/45/11/113001} {\bibfield  {journal} {\bibinfo  {journal} {Journal of Physics D Applied Physics}\ }\textbf {\bibinfo {volume} {45}},\ \bibinfo {pages} {113001} (\bibinfo {year} {2012})}\BibitemShut {NoStop}%
\bibitem [{\citenamefont {Kalinikos}\ and\ \citenamefont {Slavin}(1986)}]{kalinikos1986theory}%
  \BibitemOpen
  \bibfield  {author} {\bibinfo {author} {\bibfnamefont {B.~A.}\ \bibnamefont {Kalinikos}}\ and\ \bibinfo {author} {\bibfnamefont {A.~N.}\ \bibnamefont {Slavin}},\ }\bibfield  {title} {\bibinfo {title} {Theory of dipole-exchange spin wave spectrum for ferromagnetic films with mixed exchange boundary conditions},\ }\href {https://doi.org/10.1088/0022-3719/19/35/014} {\bibfield  {journal} {\bibinfo  {journal} {Journal of Physics C: Solid State Physics}\ }\textbf {\bibinfo {volume} {19}},\ \bibinfo {pages} {7013} (\bibinfo {year} {1986})}\BibitemShut {NoStop}%
\bibitem [{\citenamefont {Baltz}\ \emph {et~al.}(2018)\citenamefont {Baltz}, \citenamefont {Manchon}, \citenamefont {Tsoi}, \citenamefont {Moriyama}, \citenamefont {Ono},\ and\ \citenamefont {Tserkovnyak}}]{v.baltz2018antiferromagnetic}%
  \BibitemOpen
  \bibfield  {author} {\bibinfo {author} {\bibfnamefont {V.}~\bibnamefont {Baltz}}, \bibinfo {author} {\bibfnamefont {A.}~\bibnamefont {Manchon}}, \bibinfo {author} {\bibfnamefont {M.}~\bibnamefont {Tsoi}}, \bibinfo {author} {\bibfnamefont {T.}~\bibnamefont {Moriyama}}, \bibinfo {author} {\bibfnamefont {T.}~\bibnamefont {Ono}},\ and\ \bibinfo {author} {\bibfnamefont {Y.}~\bibnamefont {Tserkovnyak}},\ }\bibfield  {title} {\bibinfo {title} {Antiferromagnetic spintronics},\ }\bibfield  {journal} {\bibinfo  {journal} {Reviews of Modern Physics}\ }\textbf {\bibinfo {volume} {90}},\ \href {https://doi.org/10.1103/revmodphys.90.015005} {10.1103/revmodphys.90.015005} (\bibinfo {year} {2018})\BibitemShut {NoStop}%
\bibitem [{\citenamefont {Hwang}\ and\ \citenamefont {Das~Sarma}(2007)}]{hwang2007dielectric}%
  \BibitemOpen
  \bibfield  {author} {\bibinfo {author} {\bibfnamefont {E.~H.}\ \bibnamefont {Hwang}}\ and\ \bibinfo {author} {\bibfnamefont {S.}~\bibnamefont {Das~Sarma}},\ }\bibfield  {title} {\bibinfo {title} {Dielectric function, screening, and plasmons in two-dimensional graphene},\ }\bibfield  {journal} {\bibinfo  {journal} {Physical Review B}\ }\textbf {\bibinfo {volume} {75}},\ \href {https://doi.org/10.1103/physrevb.75.205418} {10.1103/physrevb.75.205418} (\bibinfo {year} {2007})\BibitemShut {NoStop}%
\bibitem [{\citenamefont {Grigorenko}\ \emph {et~al.}(2012)\citenamefont {Grigorenko}, \citenamefont {Polini},\ and\ \citenamefont {Novoselov}}]{a.n.grigorenko2012graphene}%
  \BibitemOpen
  \bibfield  {author} {\bibinfo {author} {\bibfnamefont {A.~N.}\ \bibnamefont {Grigorenko}}, \bibinfo {author} {\bibfnamefont {M.}~\bibnamefont {Polini}},\ and\ \bibinfo {author} {\bibfnamefont {K.~S.}\ \bibnamefont {Novoselov}},\ }\bibfield  {title} {\bibinfo {title} {Graphene plasmonics},\ }\href {https://doi.org/10.1038/nphoton.2012.262} {\bibfield  {journal} {\bibinfo  {journal} {Nature Photonics}\ }\textbf {\bibinfo {volume} {6}},\ \bibinfo {pages} {749} (\bibinfo {year} {2012})}\BibitemShut {NoStop}%
\bibitem [{\citenamefont {Pendry}\ \emph {et~al.}(2004)\citenamefont {Pendry}, \citenamefont {Martin-Moreno},\ and\ \citenamefont {Garcia-Vidal}}]{pendry2004mimicking}%
  \BibitemOpen
  \bibfield  {author} {\bibinfo {author} {\bibfnamefont {J.~B.}\ \bibnamefont {Pendry}}, \bibinfo {author} {\bibfnamefont {L.}~\bibnamefont {Martin-Moreno}},\ and\ \bibinfo {author} {\bibfnamefont {F.~J.}\ \bibnamefont {Garcia-Vidal}},\ }\bibfield  {title} {\bibinfo {title} {Mimicking surface plasmons with structured surfaces},\ }\href {https://doi.org/10.1126/science.1098999} {\bibfield  {journal} {\bibinfo  {journal} {Science}\ }\textbf {\bibinfo {volume} {305}},\ \bibinfo {pages} {847} (\bibinfo {year} {2004})}\BibitemShut {NoStop}%
\bibitem [{\citenamefont {Garc{\'i}a-Vidal}\ \emph {et~al.}(2022)\citenamefont {Garc{\'i}a-Vidal}, \citenamefont {Fern{\'a}ndez-Dom{\'\i}nguez}, \citenamefont {Martin-Moreno}, \citenamefont {Zhang}, \citenamefont {Tang}, \citenamefont {Peng},\ and\ \citenamefont {Cui}}]{garcia2022spoof}%
  \BibitemOpen
  \bibfield  {author} {\bibinfo {author} {\bibfnamefont {F.~J.}\ \bibnamefont {Garc{\'i}a-Vidal}}, \bibinfo {author} {\bibfnamefont {A.~I.}\ \bibnamefont {Fern{\'a}ndez-Dom{\'\i}nguez}}, \bibinfo {author} {\bibfnamefont {L.}~\bibnamefont {Martin-Moreno}}, \bibinfo {author} {\bibfnamefont {H.~C.}\ \bibnamefont {Zhang}}, \bibinfo {author} {\bibfnamefont {W.}~\bibnamefont {Tang}}, \bibinfo {author} {\bibfnamefont {R.}~\bibnamefont {Peng}},\ and\ \bibinfo {author} {\bibfnamefont {T.~J.}\ \bibnamefont {Cui}},\ }\bibfield  {title} {\bibinfo {title} {Spoof surface plasmon photonics},\ }\href {https://doi.org/10.1103/revmodphys.94.025004} {\bibfield  {journal} {\bibinfo  {journal} {Reviews of Modern Physics}\ }\textbf {\bibinfo {volume} {94}},\ \bibinfo {pages} {025004} (\bibinfo {year} {2022})}\BibitemShut {NoStop}%
\bibitem [{\citenamefont {Hibbins}\ \emph {et~al.}(2005)\citenamefont {Hibbins}, \citenamefont {Evans},\ and\ \citenamefont {Sambles}}]{hibbins2005experimental}%
  \BibitemOpen
  \bibfield  {author} {\bibinfo {author} {\bibfnamefont {A.~P.}\ \bibnamefont {Hibbins}}, \bibinfo {author} {\bibfnamefont {B.~R.}\ \bibnamefont {Evans}},\ and\ \bibinfo {author} {\bibfnamefont {J.~R.}\ \bibnamefont {Sambles}},\ }\bibfield  {title} {\bibinfo {title} {Experimental verification of designer surface plasmons},\ }\href {https://doi.org/10.1126/science.1109043} {\bibfield  {journal} {\bibinfo  {journal} {Science}\ }\textbf {\bibinfo {volume} {308}},\ \bibinfo {pages} {670} (\bibinfo {year} {2005})}\BibitemShut {NoStop}%
\bibitem [{\citenamefont {Garcia-Vidal}\ \emph {et~al.}(2005)\citenamefont {Garcia-Vidal}, \citenamefont {Martin-Moreno},\ and\ \citenamefont {Pendry}}]{garcia-vidal2005surfaces}%
  \BibitemOpen
  \bibfield  {author} {\bibinfo {author} {\bibfnamefont {F.~J.}\ \bibnamefont {Garcia-Vidal}}, \bibinfo {author} {\bibfnamefont {L.}~\bibnamefont {Martin-Moreno}},\ and\ \bibinfo {author} {\bibfnamefont {J.~B.}\ \bibnamefont {Pendry}},\ }\bibfield  {title} {\bibinfo {title} {Surfaces with holes in them: new plasmonic metamaterials},\ }\href {https://doi.org/10.1088/1464-4258/7/2/013} {\bibfield  {journal} {\bibinfo  {journal} {Journal of Optics A: Pure and Applied Optics}\ }\textbf {\bibinfo {volume} {7}},\ \bibinfo {pages} {S97} (\bibinfo {year} {2005})}\BibitemShut {NoStop}%
\bibitem [{\citenamefont {L{\"u}thi}\ \emph {et~al.}(1983)\citenamefont {L{\"u}thi}, \citenamefont {Mills},\ and\ \citenamefont {Camley}}]{luthi1983surface}%
  \BibitemOpen
  \bibfield  {author} {\bibinfo {author} {\bibfnamefont {B.}~\bibnamefont {L{\"u}thi}}, \bibinfo {author} {\bibfnamefont {D.}~\bibnamefont {Mills}},\ and\ \bibinfo {author} {\bibfnamefont {R.}~\bibnamefont {Camley}},\ }\bibfield  {title} {\bibinfo {title} {Surface spin waves in antiferromagnets},\ }\href {https://doi.org/10.1103/physrevb.28.1475} {\bibfield  {journal} {\bibinfo  {journal} {Physical Review B}\ }\textbf {\bibinfo {volume} {28}},\ \bibinfo {pages} {1475} (\bibinfo {year} {1983})}\BibitemShut {NoStop}%
\bibitem [{\citenamefont {Sanders}\ \emph {et~al.}(1981)\citenamefont {Sanders}, \citenamefont {Belanger}, \citenamefont {Motokawa}, \citenamefont {Jaccarino},\ and\ \citenamefont {Rezende}}]{sanders1981far}%
  \BibitemOpen
  \bibfield  {author} {\bibinfo {author} {\bibfnamefont {R.~W.}\ \bibnamefont {Sanders}}, \bibinfo {author} {\bibfnamefont {R.~M.}\ \bibnamefont {Belanger}}, \bibinfo {author} {\bibfnamefont {M.}~\bibnamefont {Motokawa}}, \bibinfo {author} {\bibfnamefont {V.}~\bibnamefont {Jaccarino}},\ and\ \bibinfo {author} {\bibfnamefont {S.~M.}\ \bibnamefont {Rezende}},\ }\bibfield  {title} {\bibinfo {title} {Far-infrared laser study of magnetic polaritons in fe${\mathrm{f}}_{2}$ and mn impurity mode in fe${\mathrm{f}}_{2}$: Mn},\ }\href {https://doi.org/10.1103/PhysRevB.23.1190} {\bibfield  {journal} {\bibinfo  {journal} {Phys. Rev. B}\ }\textbf {\bibinfo {volume} {23}},\ \bibinfo {pages} {1190} (\bibinfo {year} {1981})}\BibitemShut {NoStop}%
\bibitem [{\citenamefont {Schleck}\ \emph {et~al.}(2010)\citenamefont {Schleck}, \citenamefont {Nahas}, \citenamefont {Lobo}, \citenamefont {Varignon}, \citenamefont {Lepetit}, \citenamefont {Nelson},\ and\ \citenamefont {Moreira}}]{schleck2010elastic}%
  \BibitemOpen
  \bibfield  {author} {\bibinfo {author} {\bibfnamefont {R.}~\bibnamefont {Schleck}}, \bibinfo {author} {\bibfnamefont {Y.}~\bibnamefont {Nahas}}, \bibinfo {author} {\bibfnamefont {R.~P. S.~M.}\ \bibnamefont {Lobo}}, \bibinfo {author} {\bibfnamefont {J.}~\bibnamefont {Varignon}}, \bibinfo {author} {\bibfnamefont {M.~B.}\ \bibnamefont {Lepetit}}, \bibinfo {author} {\bibfnamefont {C.~S.}\ \bibnamefont {Nelson}},\ and\ \bibinfo {author} {\bibfnamefont {R.~L.}\ \bibnamefont {Moreira}},\ }\bibfield  {title} {\bibinfo {title} {Elastic and magnetic effects on the infrared phonon spectra of ${\text{mnf}}_{2}$},\ }\href {https://doi.org/10.1103/PhysRevB.82.054412} {\bibfield  {journal} {\bibinfo  {journal} {Phys. Rev. B}\ }\textbf {\bibinfo {volume} {82}},\ \bibinfo {pages} {054412} (\bibinfo {year} {2010})}\BibitemShut {NoStop}%
\bibitem [{\citenamefont {Moriyama}\ \emph {et~al.}(2019)\citenamefont {Moriyama}, \citenamefont {Hayashi}, \citenamefont {Yamada}, \citenamefont {Shima}, \citenamefont {Ohya},\ and\ \citenamefont {Ono}}]{moriyama2019intrinsic}%
  \BibitemOpen
  \bibfield  {author} {\bibinfo {author} {\bibfnamefont {T.}~\bibnamefont {Moriyama}}, \bibinfo {author} {\bibfnamefont {K.}~\bibnamefont {Hayashi}}, \bibinfo {author} {\bibfnamefont {K.}~\bibnamefont {Yamada}}, \bibinfo {author} {\bibfnamefont {M.}~\bibnamefont {Shima}}, \bibinfo {author} {\bibfnamefont {Y.}~\bibnamefont {Ohya}},\ and\ \bibinfo {author} {\bibfnamefont {T.}~\bibnamefont {Ono}},\ }\bibfield  {title} {\bibinfo {title} {Intrinsic and extrinsic antiferromagnetic damping in nio},\ }\bibfield  {journal} {\bibinfo  {journal} {Physical Review Materials}\ }\textbf {\bibinfo {volume} {3}},\ \href {https://doi.org/10.1103/physrevmaterials.3.051402} {10.1103/physrevmaterials.3.051402} (\bibinfo {year} {2019})\BibitemShut {NoStop}%
\bibitem [{\citenamefont {Rao}\ and\ \citenamefont {Smakula}(1965)}]{rao1965dielectric}%
  \BibitemOpen
  \bibfield  {author} {\bibinfo {author} {\bibfnamefont {K.~V.}\ \bibnamefont {Rao}}\ and\ \bibinfo {author} {\bibfnamefont {A.}~\bibnamefont {Smakula}},\ }\bibfield  {title} {\bibinfo {title} {Dielectric properties of cobalt oxide, nickel oxide, and their mixed crystals},\ }\href {https://doi.org/10.1063/1.1714397} {\bibfield  {journal} {\bibinfo  {journal} {Journal of Applied Physics}\ }\textbf {\bibinfo {volume} {36}},\ \bibinfo {pages} {2031} (\bibinfo {year} {1965})}\BibitemShut {NoStop}%
\bibitem [{\citenamefont {Rusina}\ \emph {et~al.}(2010)\citenamefont {Rusina}, \citenamefont {Durach},\ and\ \citenamefont {Stockman}}]{rusina2010theory}%
  \BibitemOpen
  \bibfield  {author} {\bibinfo {author} {\bibfnamefont {A.}~\bibnamefont {Rusina}}, \bibinfo {author} {\bibfnamefont {M.}~\bibnamefont {Durach}},\ and\ \bibinfo {author} {\bibfnamefont {M.~I.}\ \bibnamefont {Stockman}},\ }\bibfield  {title} {\bibinfo {title} {Theory of spoof plasmons in real metals},\ }\href {https://doi.org/10.1364/pmeta_plas.2010.mmd5} {\bibfield  {journal} {\bibinfo  {journal} {Applied Physics A}\ }\textbf {\bibinfo {volume} {100}},\ \bibinfo {pages} {375} (\bibinfo {year} {2010})}\BibitemShut {NoStop}%
\bibitem [{\citenamefont {Ordal}\ \emph {et~al.}(1983)\citenamefont {Ordal}, \citenamefont {Long}, \citenamefont {Bell}, \citenamefont {Bell}, \citenamefont {Bell}, \citenamefont {Alexander},\ and\ \citenamefont {Ward}}]{ordal1983optical}%
  \BibitemOpen
  \bibfield  {author} {\bibinfo {author} {\bibfnamefont {M.~A.}\ \bibnamefont {Ordal}}, \bibinfo {author} {\bibfnamefont {L.~L.}\ \bibnamefont {Long}}, \bibinfo {author} {\bibfnamefont {R.~J.}\ \bibnamefont {Bell}}, \bibinfo {author} {\bibfnamefont {S.~E.}\ \bibnamefont {Bell}}, \bibinfo {author} {\bibfnamefont {R.~R.}\ \bibnamefont {Bell}}, \bibinfo {author} {\bibfnamefont {R.~W.}\ \bibnamefont {Alexander}},\ and\ \bibinfo {author} {\bibfnamefont {C.~A.}\ \bibnamefont {Ward}},\ }\bibfield  {title} {\bibinfo {title} {Optical properties of the metals al, co, cu, au, fe, pb, ni, pd, pt, ag, ti, and w in the infrared and far infrared},\ }\href {https://doi.org/10.1364/ao.22.001099} {\bibfield  {journal} {\bibinfo  {journal} {Applied Optics}\ }\textbf {\bibinfo {volume} {22}},\ \bibinfo {pages} {1099} (\bibinfo {year} {1983})}\BibitemShut {NoStop}%
\bibitem [{\citenamefont {Dumelow}\ and\ \citenamefont {Oliveros}(1997)}]{dumelow1997continuum}%
  \BibitemOpen
  \bibfield  {author} {\bibinfo {author} {\bibfnamefont {T.}~\bibnamefont {Dumelow}}\ and\ \bibinfo {author} {\bibfnamefont {M.~C.}\ \bibnamefont {Oliveros}},\ }\bibfield  {title} {\bibinfo {title} {Continuum model of confined magnon polaritons in superlattices of antiferromagnets},\ }\href {https://doi.org/10.1103/physrevb.55.994} {\bibfield  {journal} {\bibinfo  {journal} {Physical Review B}\ }\textbf {\bibinfo {volume} {55}},\ \bibinfo {pages} {994} (\bibinfo {year} {1997})}\BibitemShut {NoStop}%
\bibitem [{\citenamefont {Kotthaus}\ and\ \citenamefont {Jaccarino}(1972)}]{kotthaus1972antiferromagnetic}%
  \BibitemOpen
  \bibfield  {author} {\bibinfo {author} {\bibfnamefont {J.}~\bibnamefont {Kotthaus}}\ and\ \bibinfo {author} {\bibfnamefont {V.}~\bibnamefont {Jaccarino}},\ }\bibfield  {title} {\bibinfo {title} {Antiferromagnetic-resonance linewidths in mn f 2},\ }\href {https://doi.org/10.1103/PhysRevLett.28.1649} {\bibfield  {journal} {\bibinfo  {journal} {Physical Review Letters}\ }\textbf {\bibinfo {volume} {28}},\ \bibinfo {pages} {1649} (\bibinfo {year} {1972})}\BibitemShut {NoStop}%
\end{thebibliography}%

\end{document}